\documentclass[]{jfm}

\usepackage{graphicx}
\usepackage{newtxtext}
\usepackage{newtxmath}
\usepackage{natbib}
\usepackage{hyperref}
\hypersetup{
    colorlinks = true,
    urlcolor   = blue,
    citecolor  = black,
}
\usepackage{epstopdf, epsfig}
\usepackage{amsmath}
\usepackage[table]{xcolor}
\usepackage{subcaption}
\usepackage{tabularx}
\usepackage{soul}
\usepackage{makecell}
\usepackage{multirow}

\newcommand{\RomanNumeralCaps}[1]

\title{On the association of secondary hairpin growth and surface pressure gradient for oscillating foils}

\author{Suyash Verma\aff{1}, Muhammad Saif Ullah Khalid\aff{1,2} and
	Arman Hemmati\aff{1}\corresp{\email{arman.hemmati@ualberta.ca}}}

\affiliation{\aff{1}Department of Mechanical Engineering, University of Alberta, Edmonton, AB T6G 2R3, Canada
\aff{2}Department of Mechanical {and Mechatronics} Engineering, Lakehead University, Thunder Bay, ON P7B 5E1, Canada}

\begin{document}
\maketitle

\begin{abstract}

The correspondence of secondary spanwise structures and pressure gradient is numerically evaluated for {a} foil, {performing heaving and pitching motion,} at a range of phase offsets (90$^\circ$ $\le \phi \le$ 270$^\circ$) and reduced frequency (0.32 $\le St_c \le$ 0.56). The Reynolds number is $Re =$ 8000. The wake is shown to be dominated by secondary hairpin-like structures that are formed due to an elliptic instability prompted by the paired primary and secondary leading edge vortex ($LEV$). The weaker secondary $LEV$ undergoes a core deformation, resulting in streamwise vorticity outflux across the {span of the foil}, and hence{,} the growth of hairpin-like structures. Evaluating pressure gradients on {the surface of the} foil reveals a unique fundamental measure to quantitatively characterize the growth of these coherent structures. Their dominant presence can be directly linked to the growth of {the} secondary $LEV$ formed due to the large{-}scale interactions under localized adverse pressure gradients. These promote a streamwise flow compression in {neighboring regions of the} primary $LEV$. This association also presents a vivid consistency across a range of kinematics. Therefore, this correspondence provides a novel procedure to investigate the mechanisms {involved in the formation} of secondary {structures} in the wake of {an} oscillating foil. 

\end{abstract}



\section{Introduction}
\label{sec:Introduction}

{Formation of vortices and evolution of complex wakes} behind an oscillating foil has been the focus of many researchers in {the} fluid dynamics community. Their studies have enriched our understanding of efficient propulsive locomotion and flow control mechanisms employed by biological swimmers and flyers \citep{Anderson1998,Triantafyllou2005,Smits2019}. The influence of complex kinematics{,} such as coupled heaving and pitching oscillation{,} on the vortex enhancement and diffusion in the wake {requires} more detailed investigation \citep{Verma2021}. These foils are largely considered as propulsors, bio-mimicking the motion of underwater swimmers and aerial flyers \citep{VanBuren2018_1}. The mechanisms of $LEV$ evolution in the wake of marine swimmers (e.g. batoid fish) and micro flyers are critical in developing propulsive {techniques} that relate to the reduction of noise propagation \citep{ORTEGA2003}. The associated instabilities of $LEV$s are further exploited to promote turbulence and early diffusion of large scale coherent structures \citep{Leweke2016}. Such mechanisms play a critical role in the advancement of technologies that particularly reduce wake hazards behind aircraft \citep{Moriche2016,Leweke2016}. 

Recent studies have also established an association between {the} growth of secondary hairpin-like vortex structures and kinematics of foils{, undergoing} combined heaving and pitching motion \citep{Verma2021,Verma2023}. The transition from heave to pitch dominated kinematics coincides with changes in {the} flow mechanism that {contributes} towards the growth of hairpin-like secondary structures, or their absence \citep{Verma2023}. At increased chord- and amplitude-based Strouhal numbers ($St_c$ and $St_A$), the stronger deformation of primary $LEVs$ {were} also discussed {previously} \citep{Chiereghin2020,Verma2021,Son2022}, although an association with secondary structures is not entirely clear. In this study, we advance the fundamental knowledge {about} the formation of secondary hairpin-like structures at a range of kinematic {parameters} by evaluating the distribution of streamwise surface pressure gradients on an oscillating foil. 

\cite{Verma2021} {provided} a comprehensive three-dimensional wake evolution {process for} an infinitely span foil{ oscillating} with a coupled heaving and pitching motion. The distribution of {the} coefficient of pressure ($C_p$) at low and high $St_c$ reveal the onset of elliptic instability mechanism for {pairs of counter-rotating vortices of unequal strength} \citep{Leweke1998,Ortega2001}. At high $St_c$, dominant secondary hairpin-like structures are observed, whose origins are {considered} to be associated with core vorticity outflux \citep{Mittal1995} of dipole rollers shed in the wake. Recently, \cite{Verma2023} reported novel findings {related to} the growth mechanisms and presence of secondary hairpin-like vortex arrangement at a constant $St_c=$ 0.32. The phase offset ($\phi$) between heaving and pitching motion {varied} in the range of 90$^\circ$ to 270$^\circ$. {The profiles of} $C_p$ reveal the presence of a paired primary and secondary $LEV$ roller arrangement{, leading} to an elliptic instability mechanism \citep{Leweke2016}, and thereby promote an outflux of core vorticity from the weaker secondary $LEV$ \citep{Verma2023}. This subsequently {leads} to thin streamwise vorticity filaments, which ultimately extend to form an arrangement of hairpin-like secondary structure in the wake \citep{Mittal1995}. Similar assessments that employed pressure measurements on wings with {finite aspect ratios} \citep{Hammer2022} are largely focused on the dominant spanwise instability characteristics, e.g.{,} wavelength of undulating $LEV$s prior to their separation from wings. The association of pressure distribution and the growth of secondary hairpin-like structures, however, remains unknown in the current literature. 

In this study, we {explain} a fundamental association and quantified links between streamwise gradients of pressure, calculated on the {boundary of the foil}, and the formation of secondary hairpin-like vortical structures{,} which were recently discussed by \cite{Verma2023}. The findings are extended {to} a range of kinematics in order to ensure a wider applicability of the association between evolution of secondary wake structures and pressure gradients on an oscillating foil. 


\subsection{Problem Description}
\label{sec:Problem_Method}

\begin{figure}
	\centering
    \includegraphics[width=10.76cm,height=5.36cm]{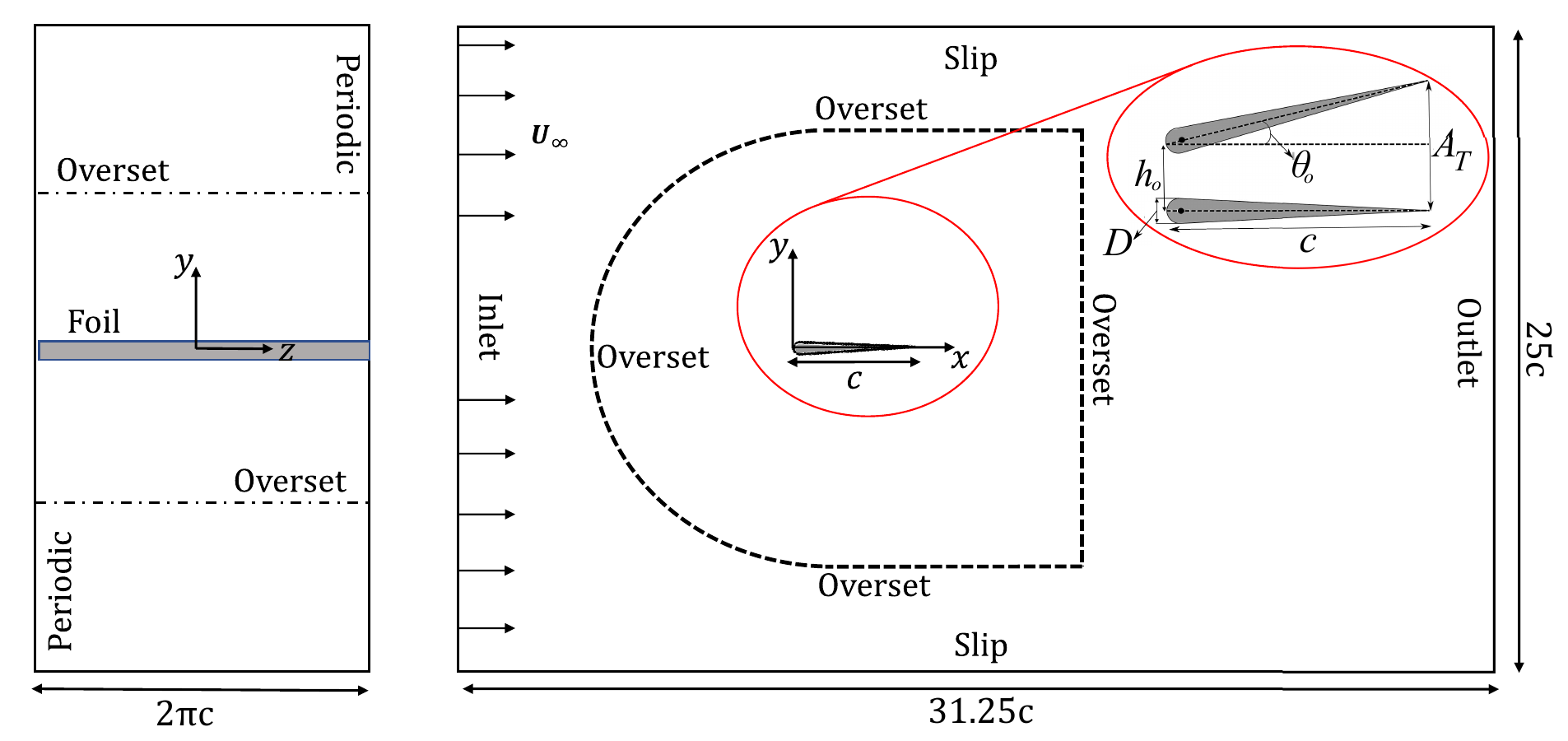}
	\caption{Schematic of the foil geometry and motion.}
	\label{fig:foil_geom}
\end{figure}

The flow around an infinitely span (2D) foil with a maximum thickness ($D$) to chord length ($c$) ratio of $D/c=0.1$ is examined numerically for a range of chord{-}based ($St_{c} = fc/U_{\infty} = 0.32 - 0.56$) and amplitude{-}based Strouhal numbers (0.05 $\le St_{A} \le$ 0.4). \cite{Andersen2017} {indicated} that significant transitions in the wake of flapping foils {were} observable at 0.2 $< St_{A} <$ 0.4. {It} also coincides with the range corresponding to the optimal propulsive efficiency in swimming mammals \citep{Triantafyllou2005,Smits2019}. The {cross section of the} foil shown in Figure \ref{fig:foil_geom} resembles a teardrop hydrofoil shape, which was used in recent experimental investigations \citep{Floryan2017,VanBuren2018_1}. 
The Reynolds number is $Re = U_{\infty} c/\nu =$ 8000, which is consistent with previous studies in this area \citep{Verma2021,Verma2023}, and agrees closely with the biological characteristics of swimming fish \citep{Anderson1998,Smits2019}. Here, $U_\infty$ and $\nu$ represent the freestream velocity and  kinematic viscosity {of the fluid}, respectively. 

The kinematics {of the foil} is {prescribed} by a coupled heaving and pitching motion, where the pitch axis is {located at} approximately 0.05$c$ from the leading edge. Figure \ref{fig:foil_geom} marks the heave and pitch amplitudes as $h_o$ and $\theta_o$, respectively. The resultant trailing edge amplitude is also shown as $A_T$. The motion profiles of heave ($h$) and pitch ($\theta$), where pitching has a phase advancement (or offset) of $\phi$ relative to heaving, are represented as $h(t)=h_{o} \sin (2 \pi f t)$ and $\theta(t)=\theta_{o} \sin (2 \pi f t+\phi)$, respectively.

%

In order to present a broader association of secondary wake structures (hairpin-like), surface pressure gradient, and kinematics {of the foil}, we also vary the phase offset ($\phi$) between heaving and pitching motion in the range of 90$^\circ$ and 270$^\circ$. {This} leads to changes in $A_T$ relative to a fixed $h_o/c$ ( = 0.25) and $\theta$ = (10$^\circ$) \citep{Verma2022_Dipole}. 
%
It is important to note that the most propulsively efficient phase offset corresponds to $\phi=$ 270$^\circ$ in our study, following the reference coordinate system employed by \cite{VanBuren2018_1}.

\subsection{Computational Method}

The continuity 
and Navier-Stokes equations 
are solved directly using OpenFOAM, which is a numerical package based on the finite-volume method. This platform is extensively used for simulating wake dynamics behind oscillating foils and panels \citep{Senturk2019,Verma2021}. {Kinematics of} the oscillatory foil is modeled using Overset Grid Assembly (OGA) method, based on a stationary background grid and a moving overset grid that are merged for the simulation \citep{Petra}. More details of the method can be found in \cite{Verma2020_Computation,Verma2021,Verma2022_Dipole}. 

The computational domain is also presented in Figure \ref{fig:foil_geom}, which highlights the C-type overset boundary containing the foil. The boundary conditions at the inlet are prescribed a uniform fixed velocity (Dirichlet) and a zero normal gradient (Neumann) for pressure. At the outlet, a zero-gradient outflow boundary condition is implied \citep{Deng2015}. The top and bottom walls are further prescribed a slip boundary condition that effectively model open-channel or free-surface flows, and closely resemble the experimental and computational conditions of \cite{VanBuren2018_1} and \cite{Hemmati2019}, respectively. At the boundary {of the foil}, a no-slip condition for velocity and a zero-gradient condition for pressure is ensured. The periodic boundary condition is further implemented on the side boundaries, coinciding with the spanwise extent of the foil. {It} provides an effective way to model {flows over bodies with infinite spans} without the end or tip effects. 

 A spatial convergence analysis is completed at $Re=$8000, $h_{o}/c=$0.25, $\theta_{o}=$15$^\circ$, $\phi=$270$^\circ$ and $St_c=$0.67. This enables comparative evaluation of the numerical results with respect to experiments of \cite{VanBuren2018_1}. Table \ref{tab:grid_verification} {summarizes} the grid convergence results involving three grids, Grid1, Grid2 and Grid3. The ratio ($\delta^*$) of minimum grid size element ($\Delta x$) to Kolmogorov scale ($\eta$) is kept approximately below 10 within the critical region near the foil ($x <$ 2.5$c$), specifically for Grid2 and Grid3 (see Table \ref{tab:grid_verification}). This region corresponds to the origin of spanwise instability and secondary structures that emerge and grow with the wake evolution \citep{Verma2021,Verma2023}. The relative error in prediction of $\overline{C_{T}}$ ($\epsilon_T=|\overline{C_{T}}_{,exp}-\overline{C_{T}}| / \overline{C_{T}}_{,exp}$), calculated with respect to the experimental results of  \cite{VanBuren2018_1}, is below 5$\%$ for Grid2. Similarly, $\epsilon_{L}^{rms} (=|C_{L,Grid3}^{rms}-C_{L}^{rms}| / C_{L,Grid3}^{rms})$, calculated with respect to the finest grid (Grid3), is below 0.1$\%$. The corresponding experimental results for $C_L$ are not yet available. This agreement in results provide sufficient confidence in Grid2 for our analysis. For more details on grid convergence, readers are referred to \cite{Verma2023}. Details for verification and validation of the numerical solver, with respect to the domain size, spatial and temporal grid, OGA solver, and boundary conditions, can be found in \cite{Hemmati2019}, \cite{Verma2020_Computation}, \cite{Verma2021} and \cite{Verma2023}. 
 
 \begin{table}
 	\caption{Grid refinement details for the current study. $N_{total}$ represents the sum of hexahedral elements in background grid and overset grid.}
 	\begin{center}
 		\def~{\hphantom{0}}
 		\begin{tabular}{lccccc||cccc}
 			Study  & $N_{total}$ & $\overline{C_{T}}$  & $C_{L}^{rms}$ & $\epsilon_T$ & $\epsilon_{L}^{rms}$ & $x=$ & $1c$ &  $2.5c$  &  $5c$\\[3pt]
 			\hline
 			Grid1 & $8.4\times10^6$ &  0.64 & 2.86 & 0.084 & 0.010 & $\delta^*=$ & 7.3 &  16.9 & 55.2 \\
 			Grid2  & $1.7\times10^7$ & 0.60 & 2.84 & 0.017 & 0.003 & $\delta^*=$ & 3.7  & 8.5 & 15.8  \\
 			Grid3  & $3.1\times10^7$ & 0.59 & 2.83 & 0.001 & - & $\delta^*=$ & 1.8 & 4.2 & 7.9  \\
 			Exp.  & - & 0.59 & - & - & - & $\delta^*=$ & - & - & -  \\
 		\end{tabular}
 		\label{tab:grid_verification}
 	\end{center}
 \end{table}

The simulations are completed using Cedar and Narval high performance clusters, operated by Digital Research Alliance of Canada. The parallel decomposition and assignment of computational domain utilizes 96 CPUs with a total of 190 GB memory and 1440 simulation hours per case.

\section{Results $\&$ Discussion}
\label{sec:Results}

We begin with a brief discussion of the growth of secondary hairpin-like structures over the span {of the foil for} a range of {kinematic conditions of the} oscillating foil. {It} is followed by a detailed discussion of streamwise pressure gradients on the boundary {of the foil} and its association with the primary and secondary $LEV$ structures. The implications of varying $St_c$ and $\phi$ on this novel finding is further highlighted, which provides sufficient confidence in employing streamwise pressure gradient as a unique quantitative measure to explain the evolution of secondary hairpin-like formations in the wake of an oscillating foil.

\subsection{Formation of secondary hairpin-like structures}

\cite{Verma2023} recently discussed the mechanisms that contribute to the growth of secondary hairpin-like structures. Here, we provide only a brief description of the wake dynamics explored by \cite{Verma2023} to set the stage for our main analysis. At a range of increasing $\phi$ (90$^\circ$$\le \phi \le$ 270$^\circ$), a paired primary and secondary $LEV$ with unequal strengths are formed at $St_c=$ 0.32, which lead to an elliptic instability of vortex cores \citep{Leweke1998,Leweke2016}. This subsequently promote the outflux of vorticity from the weaker secondary $LEV$. The shear straining on account of the primary $LEV$ further {leads} to the streamwise extension of thin hairpin-like filaments that subsequently form the dominant hairpin-like arrangement in the wake. As $\phi$ increases to 180$^\circ$, however, the dominant hairpin-like arrangement fails to originate from an instability of the pair of primary and secondary $LEV$. Rather, the primary $LEV$ paired with the $TEV$ promotes the elliptic instability of the vortex cores. Hairpin-like structures are subsequently formed through a deformed $TEV$ core, in contrast to the observations noted at $\phi=$ 90$^\circ$. The wake at $\phi=$ 225$^\circ$ and 270$^\circ$ lacks any secondary structure formation, which is associated with the decreased strength of corresponding primary $LEV$s \citep{Verma2023}. 

We further evaluate the formation of secondary hairpin-like structures at a similar range of $\phi$, while increasing $St_c$ from 0.32 to 0.56. The results (refer to the supplementary online material) confirm that at $St_c >$ 0.48, and the entire range of $\phi$ examined here, the wake is characterized by the formation of secondary hairpin-like structures. These follow the mechanism of elliptic core instability triggered by a pair of primary and secondary $LEV$. However, for $St_c \le$ 0.48 and at the onset of pitch-dominated kinematics (i.e. $\phi= $ 180$^\circ -$ 270$^\circ$), the wake either depicts a dominant hairpin-like formation, through the deformed $TEV$ core, or a complete absence of them. 

We expand on these wake dynamics by evaluating  a unique quantitative measure, in terms of streamwise pressure gradient that enable us to  understand and relate the mechanism of secondary hairpin-like growth with transitioning kinematics of {an} oscillating foil. 

\begin{figure}
	\centering
	\begin{minipage}{0.3\textwidth}
		\centering
		\subcaptionbox{\hspace*{-2.75em}}{%
			\hspace{-0.1in}	\includegraphics[width=4.5cm,height=3.75cm]{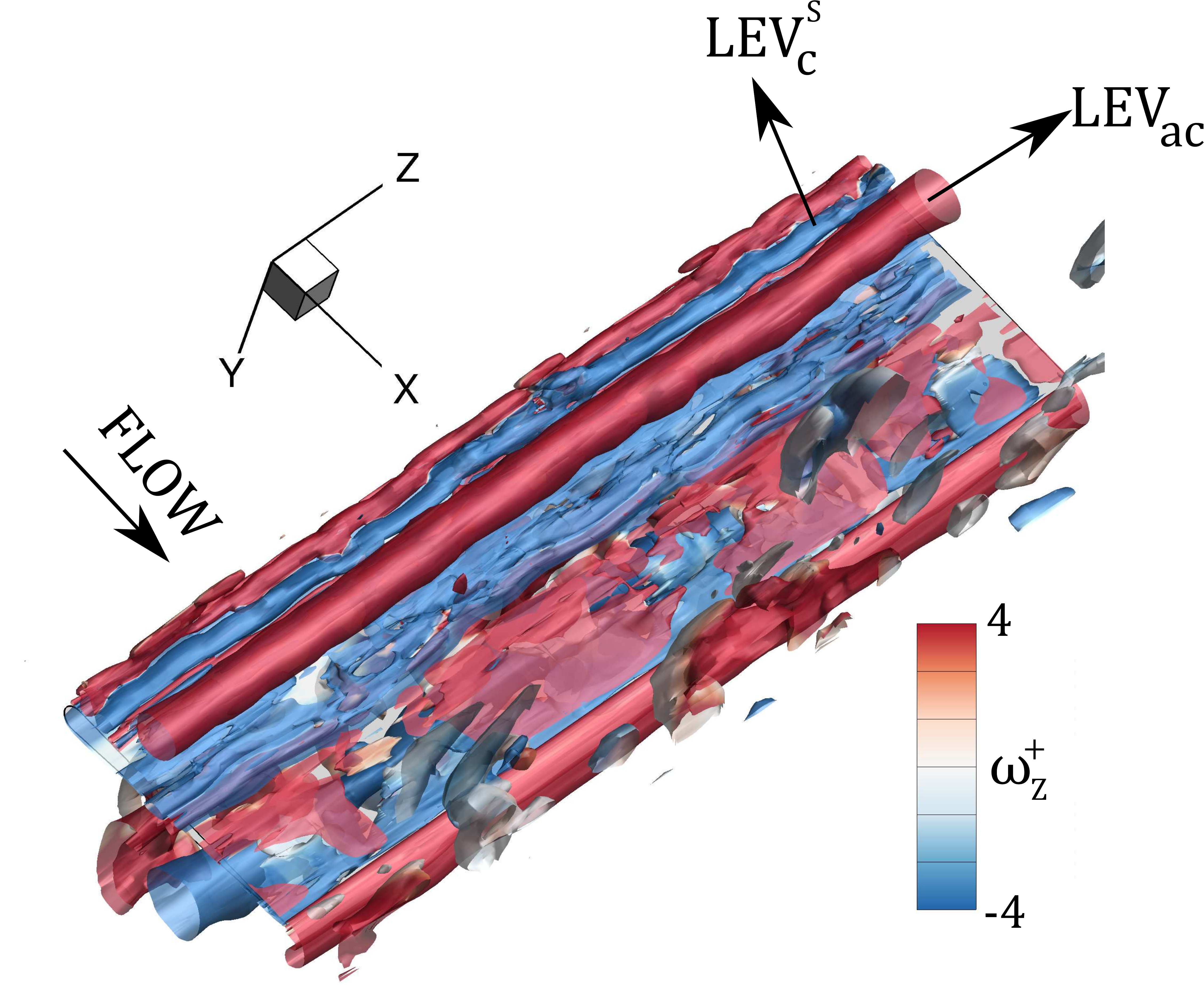}%
		}\qquad
	\end{minipage}\hfill
	\begin{minipage}{0.3\textwidth}
		\centering
		\subcaptionbox{\hspace*{-0.3em}}{%
			\hspace{-0.1in}\includegraphics[width=4.5cm,height=3.75cm]{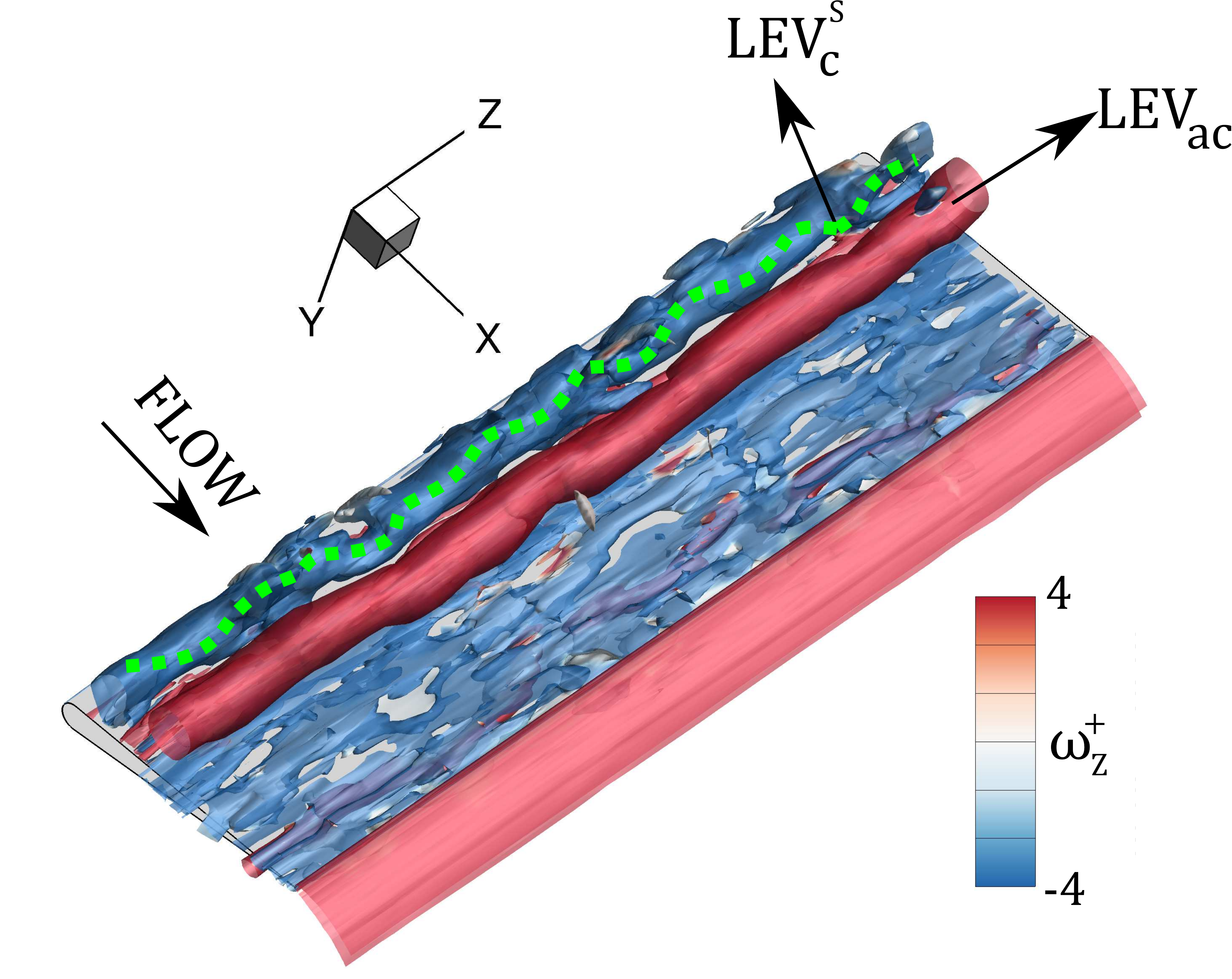}
		}\qquad
	\end{minipage}\hfill
	\begin{minipage}{0.3\textwidth}
		\centering
		\subcaptionbox{\hspace*{-2.75em}}{%
			\hspace{-0.2in}	\includegraphics[width=4.5cm,height=3.75cm]{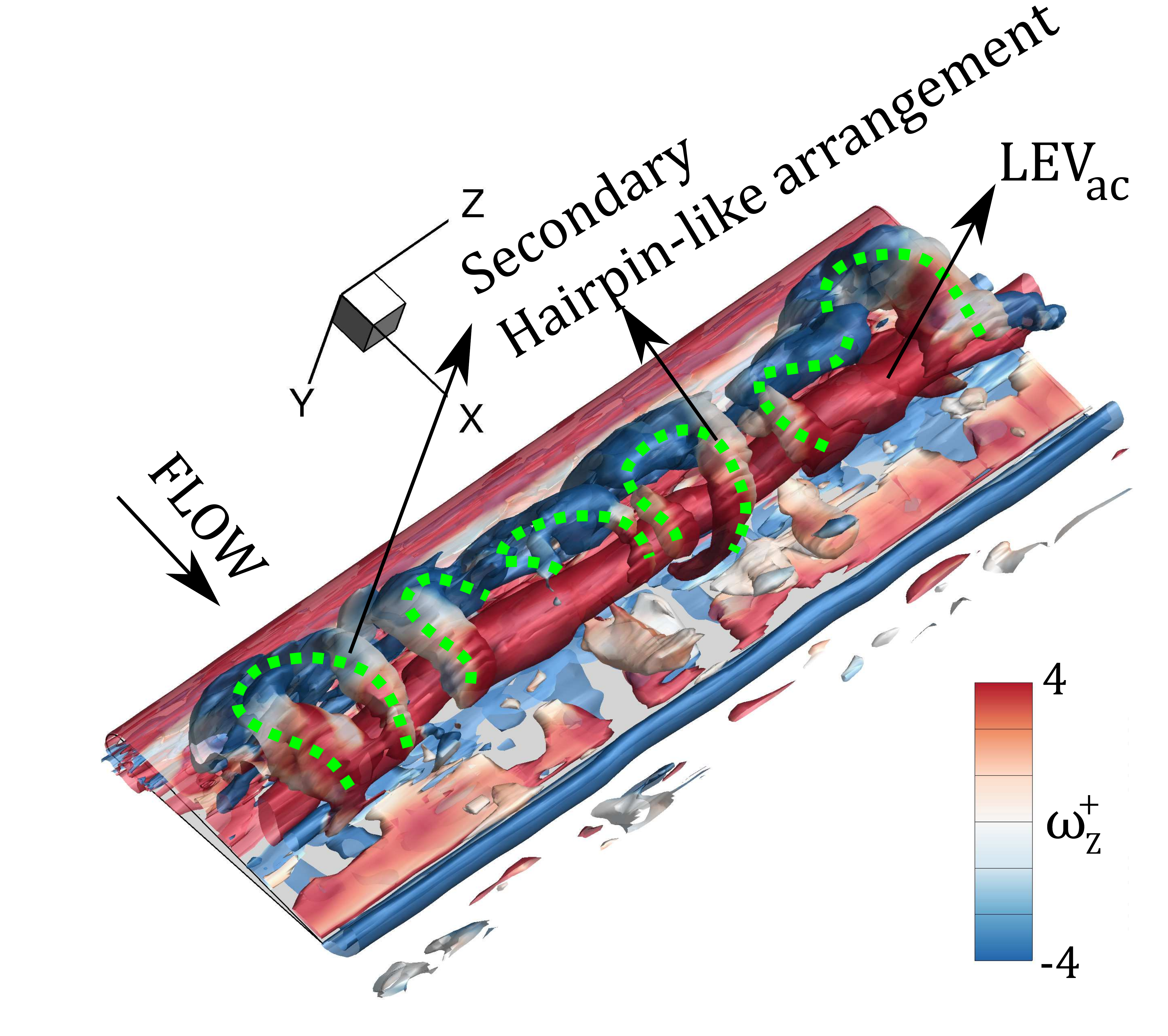}%
		}\qquad
	\end{minipage}\\
	\caption{Vortex formation (primary and secondary $LEV$) and secondary hairpin-like evolution over the foil boundary at $\phi=$ 90$^\circ$ and $St_c=$ 0.56. The time instants correspond to (a) $t^+ =$ 0, (b) $t^+ =$ 0.25 and (c) $t^+ =$ 0.5. The structures correspond to $\lambda_2^+ =$ -0.32 and are colored using spanwise vorticity ($|\omega_{z}^+| =$ 4).}
	\label{fig:Pressure_f07_90_Qualitative}
\end{figure}

\subsection{Secondary structures and streamwise pressure gradients}

We employ a span{-}averaged estimation of the coefficient of pressure ($C_p = p/(0.5\rho U_{\infty}^2)$) over the boundary {of the foil} during a half oscillation cycle. The selected kinematic setting corresponds to $\phi=$ 90$^\circ$ and $St_c =$ 0.56, although the discussion and mechanisms remain consistent for the entire parameter space {presented} in Section \ref{sec:Problem_Method}. In order to provide a qualitative outlook on the {formation of the vortices}, Figure \ref{fig:Pressure_f07_90_Qualitative}(a-c) provides an instantaneous snapshot of $LEV$s {formed} over the {surface of the foil} at first three quarters of the oscillation cycle (i.e. $t^+ =$ 0, 0.25 and 0.5). The dominant vortical structures are identified using $\lambda_2$ criterion ($\lambda_2^+ = -0.32$) following the study of \cite{Verma2023}. The arrangement of $LEV_{ac}$ and $LEV_{c}^s$ at $t^+=$ 0 reflects a pair of unequal strength (Figure \ref{fig:Pressure_f07_90_Qualitative}(a)), which subsequently triggers an elliptic instability of {the} vortex core \citep{Leweke2016}. The undulations on $LEV_{c}^s$ are enhanced at $t^+ =$ 0.25 in Figure \ref{fig:Pressure_f07_90_Qualitative}(b), owing to the developed instability. At $t^+ =$ 0.5 (Figure \ref{fig:Pressure_f07_90_Qualitative}(c)),  the growth of secondary hairpin-like structures is evident as a result of core vorticity outflux from the secondary $LEV$ \citep{Mittal1995}. The instantaneous pressure signatures{,} corresponding to the $LEV$s identified in Figure \ref{fig:Pressure_f07_90_Qualitative}, are averaged across the span and subsequently employed for calculating the streamwise pressure gradient ($dp_w/dx$).

\begin{figure}
	\centering
	\begin{minipage}{0.3\textwidth}
		\centering
		\subcaptionbox{\hspace*{-2.75em}}{%
			\hspace{-0.1in}	\includegraphics[width=4.75cm,height=3.75cm]{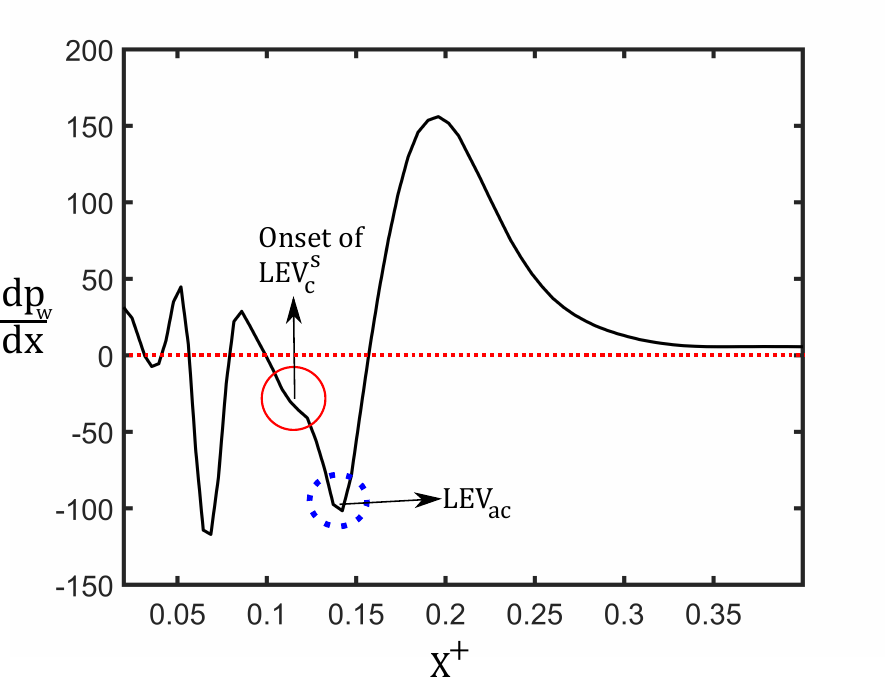}%
		}\qquad
	\end{minipage}\hfill
	\begin{minipage}{0.3\textwidth}
		\centering
		\subcaptionbox{\hspace*{-0.3em}}{%
			\hspace{-0.1in}\includegraphics[width=4.75cm,height=3.75cm]{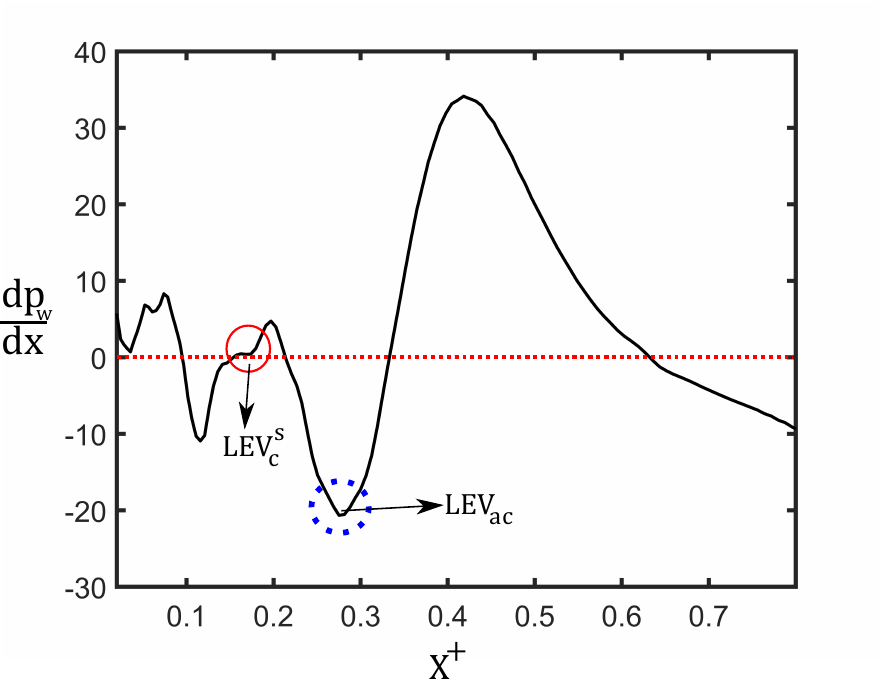}
		}\qquad
	\end{minipage}\hfill
	\begin{minipage}{0.3\textwidth}
		\centering
		\subcaptionbox{\hspace*{-2.75em}}{%
			\hspace{-0.2in}	\includegraphics[width=4.75cm,height=3.75cm]{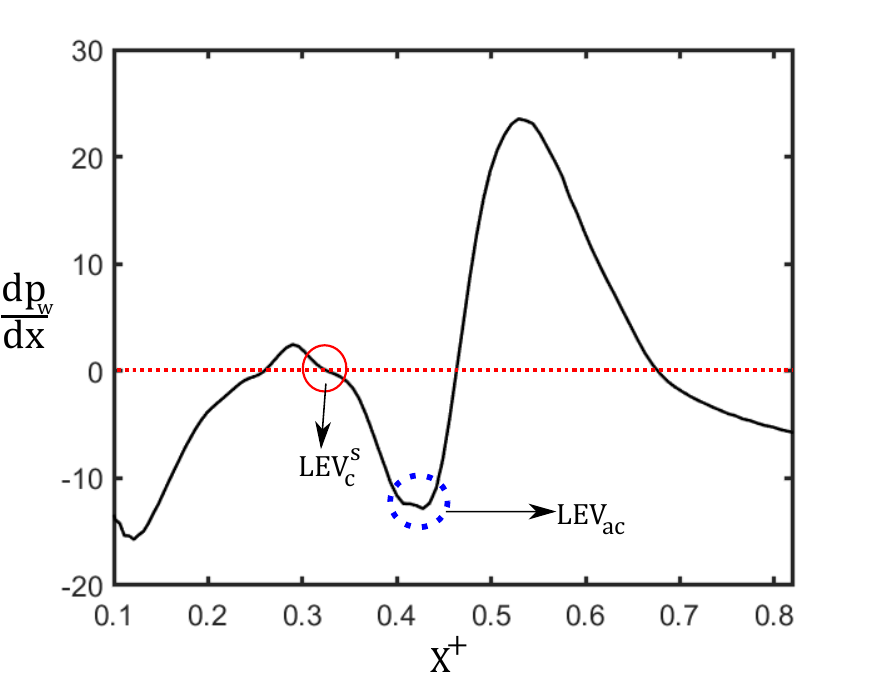}%
		}\qquad
	\end{minipage}\\
	\caption{Span averaged $dp_{w}/dx$ profiles for $\phi =$ 90$^\circ$ and $St_c =$ 0.56 $^\circ$. The time instants correspond to (a) $t^+ =$ 0, (b) $t^+ =$ 0.25 and (c) $t^+ =$ 0.5.}
	\label{fig:Pressure_f07_90}
\end{figure}

In order to evaluate an association between $dp_w/dx$ and the evolution of secondary hairpin-like structures through {the} secondary $LEV$ \citep{Verma2023}, we initially evaluate the observations relevant to the case {exhibited} in Figure \ref{fig:Pressure_f07_90_Qualitative}. Figures \ref{fig:Pressure_f07_90}(a-c) depicts the profiles of $dp_w/dx$ at first three quarter instants ($t^+ =$ 0, 0.25 and 0.5) of the oscillation cycle, where the paired primary and secondary $LEV$s are dominant. The pressure minima that coincide with the presence of {the} primary $LEV$ is highlighted in Figure \ref{fig:Pressure_f07_90}(a). The onset of {the} secondary $LEV_c^{s}$ is also marked, which is characterized by a sharp change in the profile of $dp_w/dx$. \cite{Obabko2002} provided {a} quantitative interpretation with regards to the formation of {the} secondary re-circulation zone {in the neighborhood of} a thick vortex core. Particularly, the discussion highlights {the} contribution of large{-}scale interactions in the formation of re-circulation zones. These are associated with the streamwise compression of the flow due to the existence of localized adverse pressure gradients \citep{Obabko2002}. 

To follow along the discussion of \cite{Obabko2002}, it is evident that a sudden rise from the minima of $dp_w/dx$ (coincident with $LEV_{ac}$) occurs towards a positive $dp_{w}/dx$. Hence, this region featuring a localized adverse pressure gradient is associated with the formation of $LEV_c^{s}$, similar to the observations highlighted by \cite{Obabko2002} with regards to the secondary re-circulation zone. The presence of $LEV_{c}^s$ can also be {observed} at $t^+=$ 0.25 in Figure \ref{fig:Pressure_f07_90}(b). Further ahead in the oscillation cycle at $t^+=$ 0.5, a faded signature with regards to $LEV_{c}^s$ becomes evident, which also coincides with the formation of secondary hairpin-like structures through the outflux of core vorticity from the secondary $LEV$ (see Figure \ref{fig:Pressure_f07_90_Qualitative}(c)). This is in contrast to the observations of \cite{Obabko2002}, which indicates an increase in the magnitude of drop in $dp_w/dx$, corresponding to the secondary re-circulation zone. Hence, based on the above observations, it is reasonable to discuss a plausible association between the variations of $dp_{w}/dx$ and the coincident transformation of the secondary $LEV$ to {hairpin-like vortices}. 

\begin{figure}
	\centering
			\hspace{0.0in}	\includegraphics[width=8.5cm,height=5.25cm]{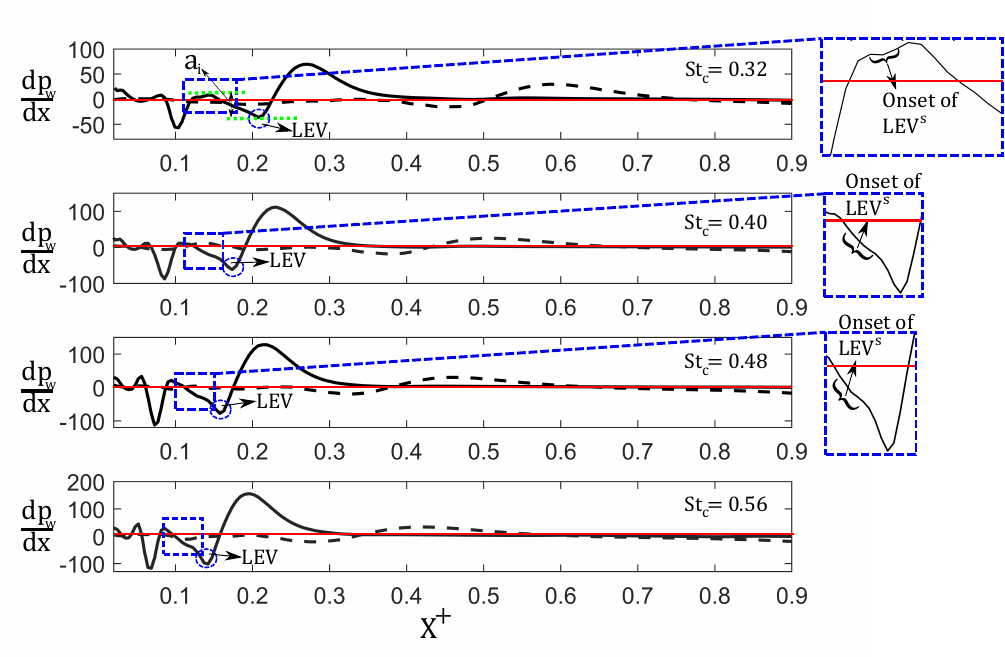}%
	\caption{Span averaged $dp_{w}/dx$ profiles for $\phi =$ 90$^\circ$ and increasing $St_c$. The straight $-$ and dashed $--$ lines corresponds to $t^+=$ 0 and $t^+=$ 0.25, respectively. 
	}
	\label{fig:Pressure_90}
\end{figure}

To address this association for a wider kinematic setting, evaluations are now  discussed for different {values of} $\phi$ and $St_c$, within the range specified in Section \ref{sec:Problem_Method}. Figure \ref{fig:Pressure_90} {exhibits} the profiles of span-averaged $dp_{w}/dx$ along the {chord of the foil} for $\phi=$ 90$^\circ$ and increasing $St_c$ from 0.32 to 0.56. The {presented} time instants correspond to $t^+ =$ 0 and 0.25, respectively. The cases that feature the growth of secondary hairpin-like structures from {the} secondary $LEV$ demonstrate a sharp change in $dp_{w}/dx$, as it increases from the local minima corresponding to the primary $LEV$. Further ahead in the cycle ($t^+\ge$ 0.25), the changes diminish as secondary hairpin-like structures grow out of {the} secondary $LEV$. For kinematics characterized by heave-domination at $\phi=$ 90$^\circ$, this characteristic association between secondary hairpin-like evolution and $dp_{w}/dx$ is consistent in the range of increasing $St_c$. 

\begin{figure}
	\centering
	\begin{minipage}{0.5\textwidth}
		\centering
		\subcaptionbox{\hspace*{-0.3em}}{%
			\hspace{-0.0in}\includegraphics[width=6.5cm,height=5.25cm]{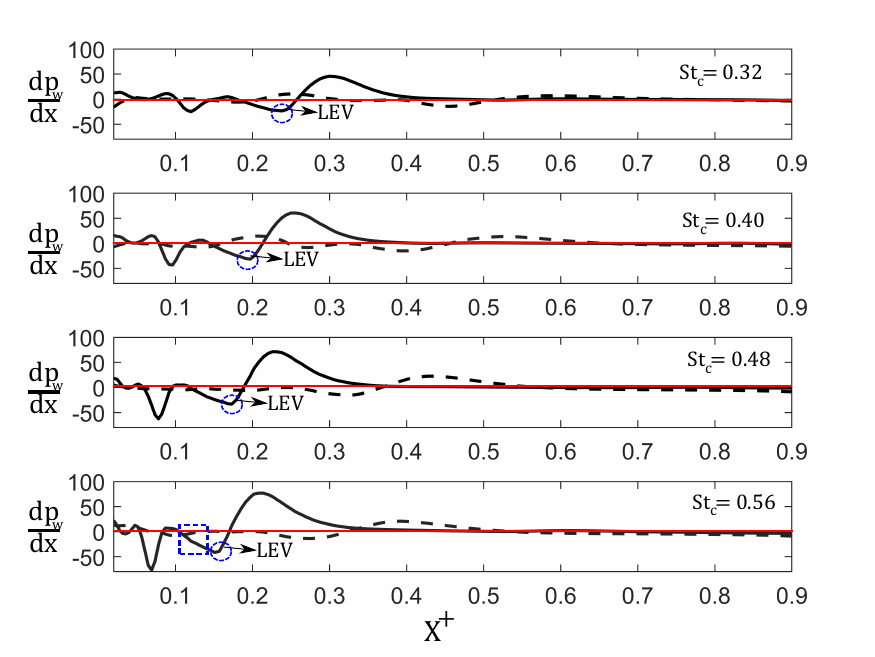}
		}\qquad
	\end{minipage}\hfill
	\begin{minipage}{0.5\textwidth}
		\centering
		\subcaptionbox{\hspace*{-2.75em}}{%
			\hspace{0.0in}	\includegraphics[width=6.5cm,height=5.25cm]{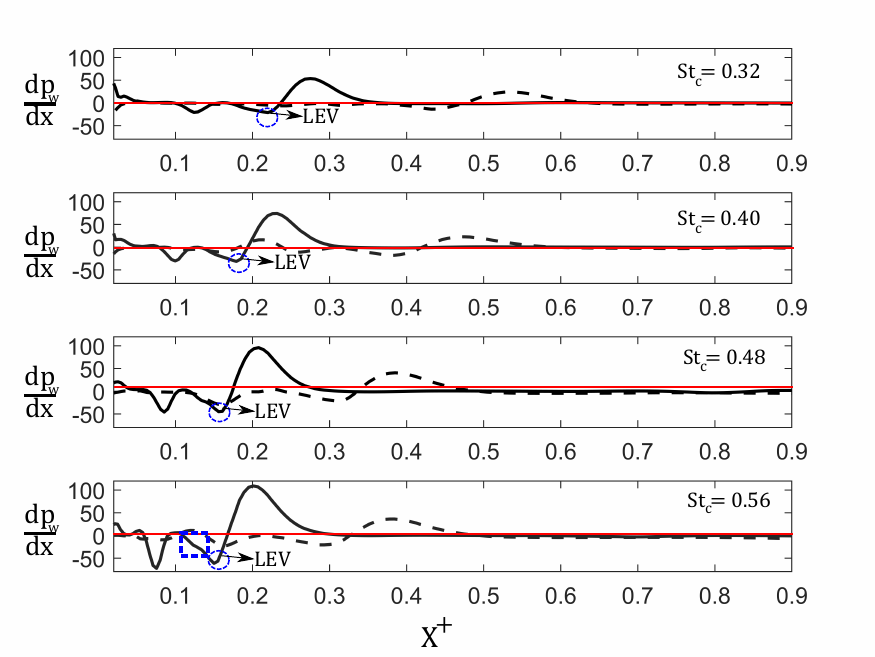}%
		}\qquad
	\end{minipage}\\
	\begin{minipage}{0.5\textwidth}
		\centering
		\subcaptionbox{\hspace*{-0.3em}}{%
			\hspace{-0.0in}\includegraphics[width=6.5cm,height=5.25cm]{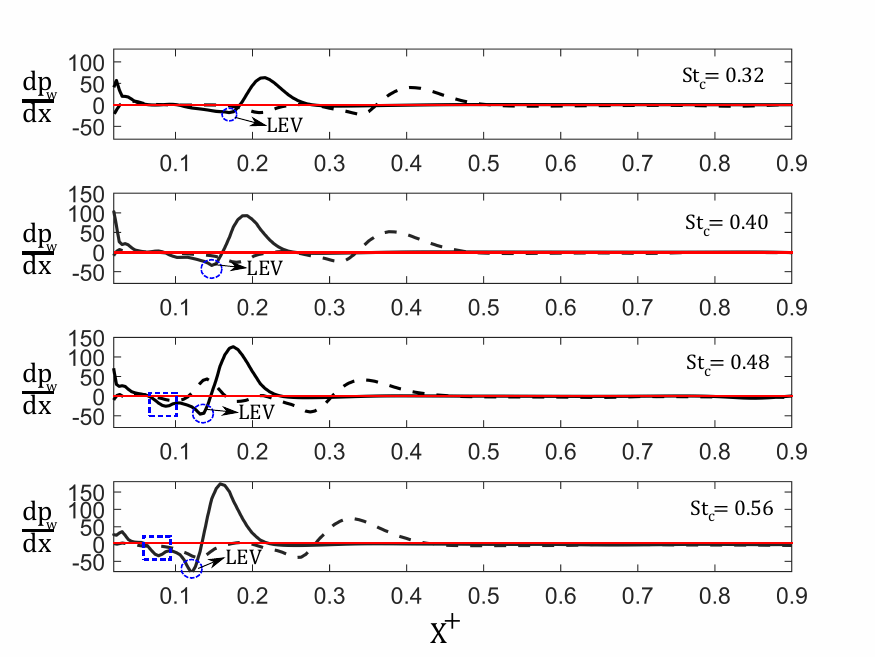}
		}\qquad
	\end{minipage}\\
	\caption{Span averaged $dp_{w}/dx$ profiles for (a) $\phi =$ 180$^\circ$, (b) $\phi =$ 225$^\circ$ and (c) $\phi =$ 270$^\circ$. The straight $-$ and dashed $--$ lines corresponds to $t^+=$ 0 and $t^+=$ 0.25, respectively.}
	\label{fig:Pressure}
\end{figure}

\begin{figure}
	\centering
	\begin{minipage}{0.2\textwidth}
		\centering
		\subcaptionbox{\hspace*{-2.75em}}{%
			\hspace{-0.1in}	\includegraphics[width=3.45cm,height=3.55cm]{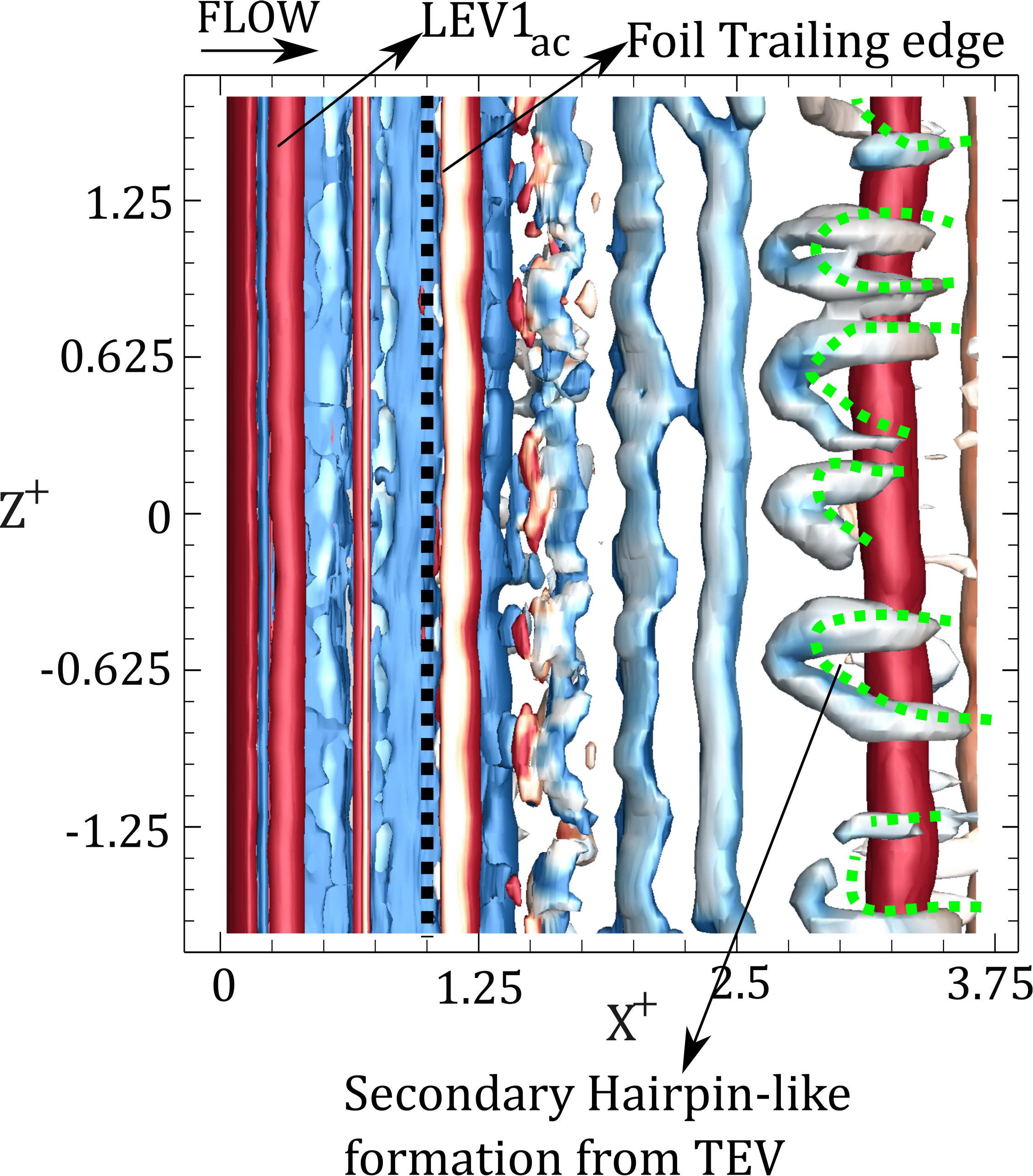}%
		}\qquad
	\end{minipage}\hfill
	\begin{minipage}{0.2\textwidth}
		\centering
		\subcaptionbox{\hspace*{-0.3em}}{%
			\hspace{-0.0in}\includegraphics[width=2.95cm,height=3.55cm]{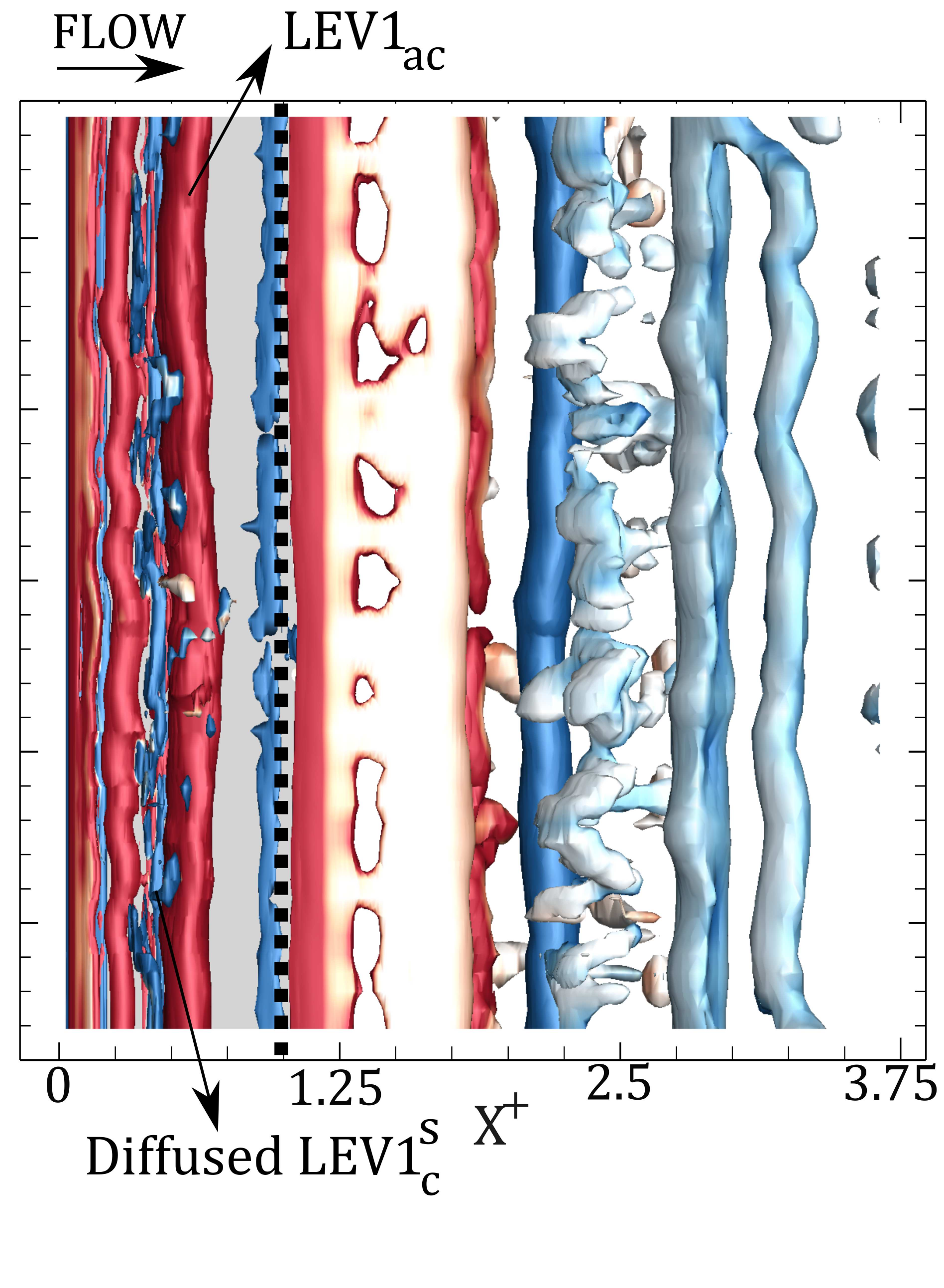}
		}\qquad
	\end{minipage}\hfill
	\begin{minipage}{0.2\textwidth}
		\centering
		\subcaptionbox{\hspace*{-2.75em}}{%
			\hspace{-0.2in}	\includegraphics[width=2.95cm,height=3.55cm]{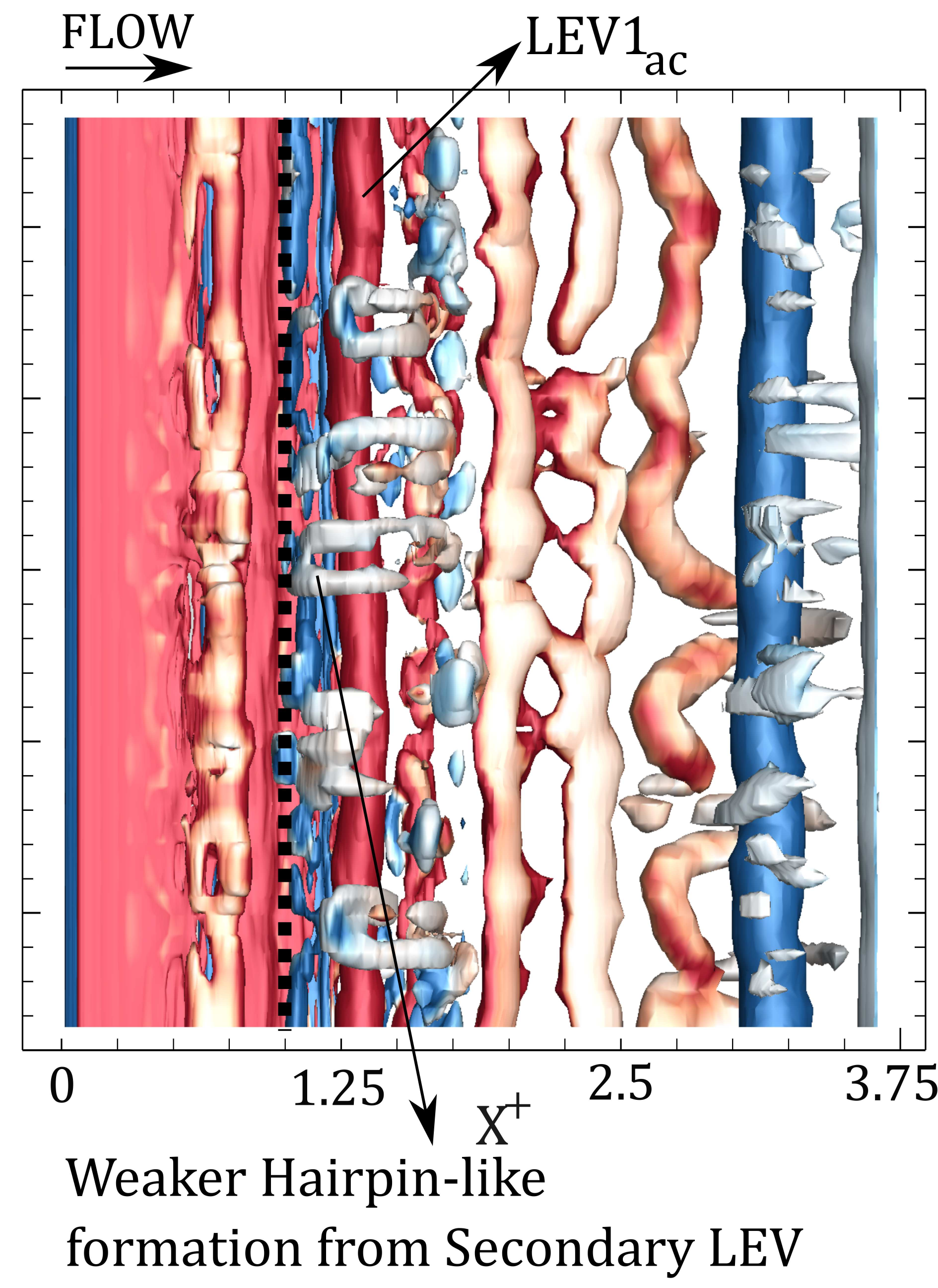}%
		}\qquad
	\end{minipage}\hfill
	\begin{minipage}{0.2\textwidth}
		\centering
		\subcaptionbox{\hspace*{-2.75em}}{%
			\hspace{-0.3in}	\includegraphics[width=2.95cm,height=3.55cm]{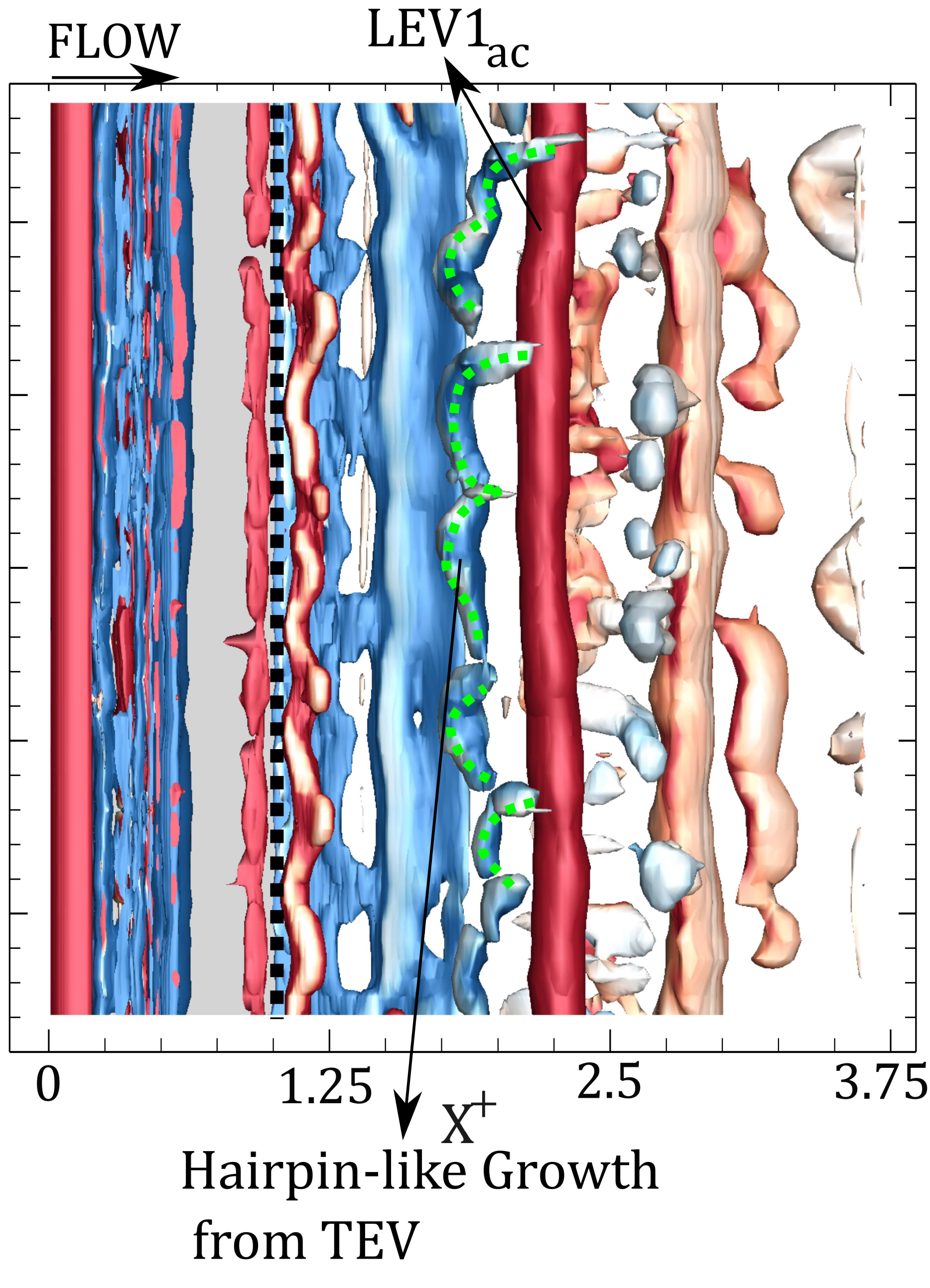}%
		}\qquad
	\end{minipage}\\
	\caption{Vortex formation (primary and secondary $LEV$) and secondary hairpin-like evolution over the foil boundary at $\phi=$ 180$^\circ$ and $St_c=$ 0.32. The time instants correspond to (a) $t^+ =$ 0, (b) $t^+ =$ 0.25, (c) $t^+ =$ 0.5 and (d) $t^+ =$ 0.75. The structures correspond to $\lambda_2^+ =$ -0.32 and are colored using spanwise vorticity ($|\omega_{z}^+| =$ 4).}
	\label{fig:Pressure_f04_180}
\end{figure}

Figure \ref{fig:Pressure}(a-c) {shows} the variations of span-averaged $dp_{w}/dx$ for $\phi=$ 180$^\circ$, 225$^\circ$ and 270$^\circ$, respectively. The observations for $\phi >$ 90$^\circ$ demonstrates sharp changes in $dp_{w}/dx$ only at $St_c \ge$ 0.48. This also coincides with the qualitative results similar to those presented in Figure \ref{fig:Pressure_f07_90_Qualitative}, which highlight the evolution of secondary hairpin-like structures from {the} secondary $LEV$ at similar kinematics. For $\phi=$ 180$^\circ$ and 225$^\circ$, observations at $St_c <$ 0.48 {suggest} that the dominant secondary hairpin-like structures are associated with {the} instability triggered by the pair of $LEV-TEV$, rather than a pair of primary and secondary $LEV$. In order to {further illustrate} this mechanism, we look at the wake visualization in Figure \ref{fig:Pressure_f04_180}(a-c). These plots qualitatively depicts the formation of dominant secondary hairpin-like structures through a $LEV-TEV$ pair at $\phi=$ 180$^\circ$ and $St_c=$ 0.32 \citep{Verma2023}. The secondary $LEV$ (marked as $LEV1_{c}^s$) at $t^+ =$ 0.25 in Figure \ref{fig:Pressure_f04_180}(b) appears much weaker and diffused, which only results in the growth of thin hairpin-like {flow structures} at $t^+=$ 0.5 (see Figure \ref{fig:Pressure_f04_180}(c)). These structures soon lose their coherence {due to} diffusion in the near wake (Figure \ref{fig:Pressure_f04_180}(d)). However, the dominant secondary hairpin-like arrangement emerged from the core vorticity outflux of $TEV${,} as seen in Figure \ref{fig:Pressure_f04_180}(d). The dominant hairpin-like arrangement in the wake in Figure \ref{fig:Pressure_f04_180}(a), are formed on account of the $LEV-TEV$ instability. We observe that the trends of increasing $dp_{w}/dx$ for such cases also appear flatter and closer to zero, which reflects a low streamwise flow compression, and thus the absence of a strong secondary $LEV$. 

\begin{figure}
	\centering
    \includegraphics[width=6.36cm,height=4.76cm]{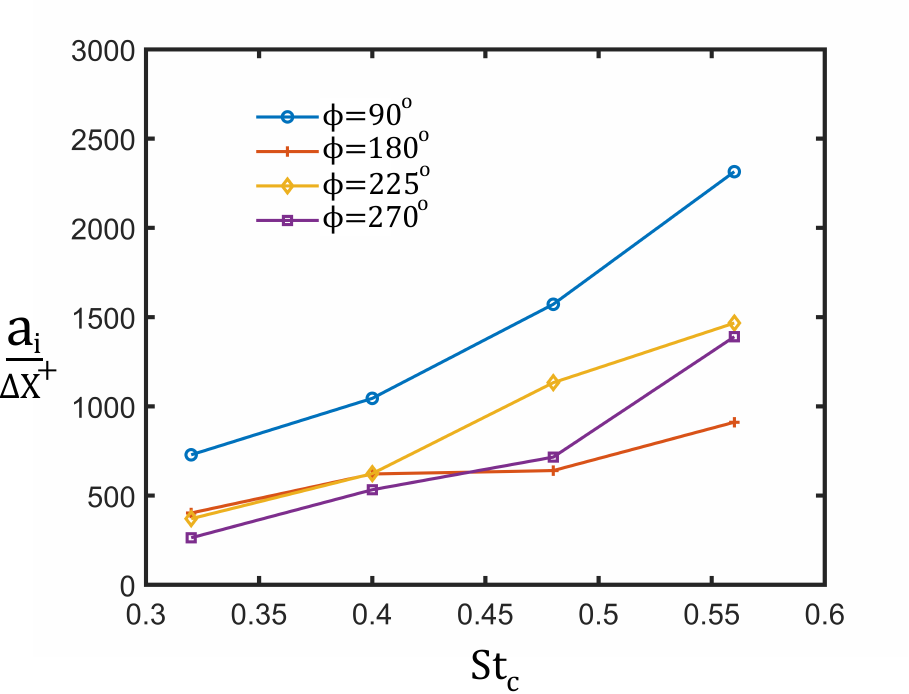}
	\caption{Variation of slope ($a_{i}/\Delta X^+$) with $St_c$ and 90$^\circ$$\le \phi \le$270$^\circ$.}
	\label{fig:Slope_Stc}
\end{figure} 

To further understand the role of $dp_{w}/dx$ in quantitative characterization of the evolution of secondary hairpin-like structures, we calculate the slope of  $dp_{w}/dx$ with respect to the streamwise distance along the foil. These estimates are further focused on a localized region that {features} the rise from a pressure minima (marked by blue dotted square in {Figures} \ref{fig:Pressure_90} and \ref{fig:Pressure}), coinciding with the primary $LEV$ structures. {It} provides an indicative measure for the increasing streamwise flow compression as $St_c$ increases, and hence expands on the association of $dp_{w}/dx$ and the growth of secondary hairpin-like structures. 
Figure \ref{fig:Slope_Stc} demonstrates the variation in the slope of $dp_{w}/dx$ for every {kinematic} setting considered in this study. Here, the slope is {computed} as the ratio of $a_{i}$ and $\Delta X^+$, where $a_{i}$ represents the magnitude of increasing $dp_{w}/dx$ (marked in Figure \ref{fig:Pressure_90}), and $\Delta X^+$ {denotes} the relative streamwise distance between the minima and local maxima following the rise of $dp_{w}/dx$. With increasing $St_c$, the slope of $dp_{w}/dx$ increases in Figure \ref{fig:Slope_Stc}. {It} suggests an environment of larger adverse pressure gradient and a stronger streamwise flow compression, 
which further coincides with a consistent secondary hairpin-like evolution from the secondary $LEV$, as $St_c$ {becomes greater than} 0.48. 

The novel association described above presents a very useful quantitative tool in characterization of the changes in mechanisms that govern the growth of secondary spanwise structures. These particularly coincide with changes in {kinematics of an} oscillating foil from heave- to onset of pitch{-}domination. For example, \citep{verma2023PRS} recently investigated and discussed such transitions, namely, Series $A$ and Series $B$. The trends for $dp_{w}/dx$ explained in this study accurately quantifies the identification of these transitions. Particularly, the increasing slope of $dp_{w}/dx$ with increasing $St_c$ coincides with the growth of secondary hairpin-like structures through a primary and secondary $LEV$ pair (i.e. Mechanism $``1"$ in \citep{verma2023PRS}), which also coincides with the Series $A$ transition at different $\phi$ \citep{verma2023PRS}.

\section{Conclusion}
\label{sec:Conclusion}

A fundamental association between the evolution of secondary structures, in the form of {a} spanwise hairpin-like arrangement, and the pressure gradient along the streamwise flow is {computationally} investigated for the case of infinitely span oscillating foils. Under large adverse gradients, a stronger streamwise flow compression promotes the  formation of {a} dominant secondary $LEV$ {in the neighborhood of} the paired primary $LEV$. The growth of {the} secondary $LEV$ {leads} to an elliptic instability mechanism \citep{Leweke2016}{,} resulting in a core vorticity outflux from the secondary $LEV$, and a subsequent growth of secondary hairpin-like structures. The highlighted association is consistent across the range of increasing $St_c$ and $\phi$, and thus reflects a unique quantitative measure of the growth of secondary wake structures behind oscillating foils.  

\backsection[Funding]{This research has received support from the Canada First Research Excellence Grant. The computational analysis was completed using Compute Canada clusters. }
%

\backsection[{\bf Declaration of Interests}] {The authors report no conflict of interest.}

%
%
%



\bibliographystyle{jfm}
\bibliography{MixMotion_Dual_Vortex}

\end{document}